\documentclass[12pt]{article}

\usepackage[totalheight = 23cm, totalwidth = 17cm]{geometry}
\usepackage{amssymb,amsmath,amsfonts,amsbsy,graphicx,bm}

\usepackage{color,cancel,ulem}

\newcommand{\dis}[1]{\begin{equation}\begin{split}#1\end{split}\end{equation}}

\begin{document}

\begin{titlepage}

\begin{center}

{\setlength{\baselineskip}{1.7\baselineskip}

{\LARGE \bf 
 Kaluza-Klein tower and bubble nucleation in   six dimensional Einstein-Maxwell theory  
}

\par}

\vskip 1.0cm

{\large
Min-Seok Seo$^{a}$ 
}

\vskip 0.5cm

{\it
$^{a}$Department of Physics Education, Korea National University of Education,
\\ 
Cheongju 28173, Republic of Korea
}

\vskip 1.2cm

\end{center}

\begin{abstract}
 
  We study the implication of the distance and the cobordism conjecture on the 6-dimensional Einstein-Maxwell theory compactified on $S^2$.
  In this toy model,  the radion potential is stabilized by the conspiracy of the curvature of $S^2$ and the flux through $S^2$ parametrized by $f$, and uplifted by the positive 6-dimensional cosmological constant parametrized by $\lambda$.
  When $\lambda=0$, the radion is stabilized at the anti-de Sitter (AdS) vacuum, which cannot be interpolated to the Minkowski vacuum since the Kaluza-Klein (KK) tower descends from UV in the vanishing limit of   the 4-dimensional cosmological constant. 
  For nonzero $\lambda$ which realizes the metastable de Sitter (dS) vacuum, as well as the AdS and the Minkowski vacuum,  such an obstruction can be found provided the combination $f^2\lambda$ is fixed and the limit $\lambda\to 0$ is taken. 
  Moreover, the 6-dimensional Einstein-Maxwell theory allows the transition between vacua through the nucleation of the bubble.
  In this case, the values of the 4-dimensional cosmological constant inside and outside the bubble   are  different as $f$ is changed at the bubble wall, while $\lambda$ remains unchanged.  
  Regarding the AdS vacuum with the vanishing curvature radius as the `nothing', we find that the transition from the metastable dS vacuum to the nothing is not prevented by the descent of the KK tower since $f^2\lambda$ is not fixed.

\end{abstract}

\end{titlepage}

\newpage

\section{Introduction}

 Whereas the values of physical parameters in the low energy effective field theory (EFT) can be measured experimentally,  we expect that the underlying fundamental theory provides a dynamical mechanism explaining how these values are determined.
 In string theory, the moduli stabilization does the job : under the compactification, parameters in the 4-dimensional EFT are controlled by the size and the shape of the extra dimensions, which are fixed by the stabilization of  scalars called moduli. 
 Typically, a large number of possibilities for choosing the flux quanta in the six extra dimensions allow the moduli potential to have a local minimum \cite{Giddings:2001yu}, the collection of which is called the string landscape \cite{Susskind:2003kw}.
 But still, in order to explain the tiny, positive cosmological constant as we observe it, the moduli need to be stabilized at the metastable de Sitter (dS) vacuum, rather than the anti-de Sitter (AdS) vacuum.
 This is achieved by introducing the uplift generated by anti-branes in addition \cite{Kachru:2002gs}.
 Several models for the metastable dS vacuum built in this way have been proposed \cite{Kachru:2003aw, Balasubramanian:2005zx}, which contributed to the active debate on whether the metastable dS vacuum can be realized in the controllable parameter regime \cite{Dine:1985he}.

  The roles of the flux and the brane in the moduli stabilization can be simply illustrated by the toy model,  the 6-dimensional Einstein-Maxwell theory compactified on $S^2$ with the magnetic flux through $S^2$ \cite{Freund:1980xh, Randjbar-Daemi:1982opc} (see \cite{Denef:2007pq} for a pedagogical review).
   In this model, the radion, the modulus determining the radius of $S^2$, is stabilized through the conspiracy of two potential terms, each of which is generated by the curvature of $S^2$ and the magnetic flux through $S^2$, respectively.
    This indeed is a typical way that the moduli are stabilized in string theory, in which the curvature term may be replaced by another flux term or the non-perturbative effect, depending on the model. 
Moreover, even though the origin is different, the positive 6-dimensional cosmological constant mimics the role of the uplift term.

  On the other hand, recent swampland program \cite{Vafa:2005ui} (for reviews, see  \cite{Brennan:2017rbf, Palti:2019pca, vanBeest:2021lhn, Grana:2021zvf, Agmon:2022thq, VanRiet:2023pnx}) suggests to revisit the moduli stabilization and see if there is a missing ingredient which makes the EFT description   inconsistent with quantum gravity.
  Many studies along this line rely on the distance conjecture \cite{Ooguri:2006in}.
 It states that the infinite distance limit of the moduli space corresponds to a corner of the landscape, at which the EFT description becomes invalid as a tower of states descends from UV.
 The 6-dimensional Einstein-Maxwell theory naturally contains a tower of states, a set of Kaluza-Klein (KK) modes, the mass scale of which becomes extremely tiny in the decompactification limit, i.e., the limit of  the infinitely large radius of $S^2$.
 Therefore, it is instructive to investigate how the radius of $S^2$, hence the KK mass scale, is determined by the  flux and the uplift in the theory in detail, focusing on the decompactification limit.
 Since the 4-dimensional cosmological constant $\Lambda_4$ is determined by the flux and the uplift  as well, we will eventually arrive at the scaling law obeyed by $\Lambda_4$ and the KK mass scale. 
Regarding this, there have been conjectures which try to argue that the tiny value of $\Lambda_4$  implies the low tower mass scale \cite{Lust:2019zwm, Palti:2020tsy, Cribiori:2021gbf, Castellano:2021yye, Blumenhagen:2022zzw, Seo:2023ssl}.
 If it were the case, our universe  having $\Lambda_4 \simeq 10^{-123}M_4^4$, where $M_4$ is the 4-dimensional Planck mass, is quite close to the corner of the landscape.
 Presumably, a relevant tower of states may be the KK tower, which implies the large size of the extra dimensions \cite{Montero:2022prj}.
Moreover, $\Lambda_4$ provides the IR mass scale of the EFT, thus  from the scaling law we can see if there is the scale separation between the IR  and the  KK mass scale  \cite{Tsimpis:2012tu, Gautason:2015tig, Shiu:2022oti}.
 The first purpose of this article is to explore the scaling law and the scale separation in  the 6-dimensional Einstein-Maxwell theory and discuss the  implications of them.

  It is also remarkable that the 6-dimensional Einstein-Maxwell theory admits the instanton solution describing the tunnelling between different flux vacua, which in particular includes the nucleation of the bubble of `nothing'.
 Here the bubble of nothing indicates the bubble containing not only no matter, but also no spacetime, which is known to exist in the presence of the  extra dimensions \cite{Witten:1981gj}.
At the wall of the bubble of nothing, the radius of the extra dimensions shrinks to  zero size, and at the same time noncompact 4-dimensional spacetime pinches off such that it cannot extend to the region inside the bubble wall.
 In the 6-dimensional Einstein-Maxwell theory, the magnetic flux, one of ingredients determining the value of $\Lambda_4$, can be discharged by the  black 2-brane forming the bubble wall \cite{Blanco-Pillado:2009lan}.
 Then two vacua separated by the bubble wall have   different values of $\Lambda_4$ due to the  different values of the flux,  while they have the same  value  of the 6-dimensional cosmological constant which generates the uplift.
 Moreover, a new bubble can be nucleated inside  the bubble, which leads to the sequential nucleation of bubbles having different values of $\Lambda_4$.
When the magnetic flux is completely discharged by the brane, i.e., the flux inside  the bubble  vanishes, the value of $\Lambda_4$ inside the  bubble becomes negative infinity, which corresponds to  the AdS vacuum with the vanishing curvature radius.  
Since the radius of $S^2$ vanishes as well, we can regard this bubble as the bubble of nothing.
 It indicates that spacetime can reach the nothing through a (series of) nucleation of the bubble(s), which is extensively studied in \cite{Brown:2010bc, Brown:2010mf, Brown:2011gt}.
 
 Recently, the bubble of nothing has drawn attention (see, for example, \cite{Draper:2021ujg, Draper:2021qtc, Buratti:2021fiv, Angius:2022aeq, Calderon-Infante:2023ler, Friedrich:2023tid, Blanco-Pillado:2023hog, Blanco-Pillado:2023aom}) in the context of the swampland program.
 For instance, the cobordism conjecture \cite{McNamara:2019rup} (see \cite{Andriot:2022mri} for a review) argues that the  compact space consistent with quantum gravity can  shrink to   zero size without any topological obstruction.
 This conjecture is closely connected to an old idea that the global symmetry is broken by quantum gravity \cite{Banks:2010zn}.
 For an illustration,  we may consider  the compact ball made of the black hole interior, which can smoothly evolve into the nothing according to the cobordism conjecture \cite{Andriot:2022mri}.
 This is nothing more than the complete evaporation of  the black hole, in which the global symmetry is not conserved since otherwise information is lost.
 If the conjecture is true, the transition to the nothing in which the internal manifold ($S^2$ in our case) shrinks to  zero size must be the topology preserving process.
 This motivates us to discuss the implication of the cobordism conjecture on the vacuum transition in the 6-dimensional Einstein-Maxwell theory.
 The vacuum transition to the nothing through the nucleation of the bubble is quite different process from that in \cite{Draper:2021ujg, Draper:2021qtc, Buratti:2021fiv, Angius:2022aeq, Calderon-Infante:2023ler, Friedrich:2023tid} where the radion is not stabilized but rolls along the potential, but this is another way that the change in the value of the radion takes place along some particular direction of the coordinate space (see \cite{Basile:2023rvm, Shiu:2023bay} for a similar discussion).

 This article is organized as follows.
 In Sec. \ref{sec:6DEM}, we review the features of  the 6-dimensional Einstein-Maxwell theory compactified on $S^2$ relevant to our discussion.
 Based on this, we consider in Sec. \ref{sec:ScaDis} the implication of the distance conjecture on the radion stabilization, presenting the relation between $\Lambda_4$ and the KK mass scale.
 Sec. \ref{sec:BoN} is devoted to the discussion on the transition between different flux vacua through the nucleation of the bubble, focusing on the transition to the nothing which is relevant to the cobordism conjecture.
 Then we conclude.
 Some details of the bubble are reviewed in App. \ref{app:wholesol}.

\section{Review on the  6-dimensional Einstein-Maxwell theory }
\label{sec:6DEM}

 The action for the 6-dimensional Einstein-Maxwell theory is  
 \dis{S_{\rm EM}=\int d^6x \sqrt{-G}\Big[\frac{M_6^4}{2}{\cal R}^{(6)}-\frac14 F_{AB}F^{AB}-\Lambda_6\Big],}
 where $M_6$ is the 6-dimensional Planck mass, $F_{AB}$ is the electromagnetic field strength ($A, B=0,1,\cdots 5$), and $\Lambda_6$ is the 6-dimensional cosmological constant of mass dimension six.
Solutions to the equations of motion describing the  compactification on $S^2$ by the magnetic flux are given by  
 \dis{&ds^2=e^{-\frac{\phi-\langle\phi\rangle}{M_4}}g_{\mu\nu}dx^\mu dx^\nu +e^{\frac{\phi-\langle\phi\rangle}{M_4}}\big(e^{\frac{\langle\phi\rangle}{M_4}}R_0^2\big) (d\psi^2+\sin^2\psi d\omega^2),
 \\
 &F_2=\frac{g_6N}{4\pi}\sin\psi d\psi\wedge d\omega,\quad\quad N \in \mathbb{Z}.\label{eq:EMsol}}
 While $R_0$, the fiducial radius of $S^2$, can be chosen arbitrarily,   the physically meaningful radius is fixed by the stabilization of the radion $\phi$ to $\langle\phi\rangle$ such that $\langle R\rangle=e^{\frac{\langle\phi\rangle}{2M_4}}R_0$, from which the squared 4-dimensional Planck mass is written as
 \dis{M_4^2=4\pi \langle R\rangle^2 M_6^4 = 4\pi e^{\frac{\langle\phi\rangle}{M_4}}R_0^2 M_6^4.\label{eq:Planck4}}
 Meanwhile, the magnetic coupling   $g_6$ has the mass dimension one as the dimensionless 4-dimensional electric coupling  is obtained by the relation $e_4^2=e_6^2/(4\pi\langle R\rangle^2)$, where $e_6$ is the 6-dimensional electric coupling, the inverse of  $g_6$.
 We also note that as a result of the compactification, the mixed  components (between the compact and the noncompact directions) of the metric as well as $A_{\mu}$ become the 4-dimensional gauge bosons.
 As shown explicitly in  \cite{Randjbar-Daemi:1982opc}, the gauge symmetry  consistent with the 4-dimensional Minkowski spacetime is given by SU(2)$\times$U(1).
 Whereas it allows another solution to the equations of motion containing the SU(2) instanton, we concentrate on the solution given by \eqref{eq:EMsol}.
 The dependence of the 6-dimensional metric on $\phi$ is determined by the requirement that the kinetic term of $\phi$ is written in the canonical form in the Einstein frame action after integrating out the extra dimensions
 \footnote{When D-dimensional spacetime is compactified on the internal manifold of dimension $n=D-4$, the metric giving the canonical kinetic term of the volume modulus $\phi$ in the Einstein frame action is written as
 \dis{ds^2=e^{-\sqrt{\frac{2n}{n+2}}\phi}g_{\mu\nu}dx^\mu dx^\nu +e^{2\sqrt{\frac{2}{n(n+2)}}\phi}g_{mn}dy^mdy^n.}
 Our case corresponds to $n=2$.
 }
 :
 \dis{S=\int d^4x\sqrt{-g}\Big[\frac{M_4^2}{2}{\cal R}^{(4)}-\frac12g^{\mu\nu}\partial_\mu \phi\partial_\nu\phi-V(\phi)\Big],\label{eq:4DEFT}}
where the radion potential $V(\phi)$ is given by 
\dis{V(\phi)=\frac{4\pi}{M_4^4}\Big[\frac{1}{32\pi^2}\frac{(g_6N)^2}{M_6^2}\frac{e^{-3\frac{\phi}{M_4}}}{R_0^6M_6^6}-\frac{e^{-2\frac{\phi}{M_4}}}{R_0^4M_6^4}+\frac{\Lambda_6}{M_6^6}\frac{e^{- \frac{\phi}{M_4}}}{R_0^2M_6^2}\Big].\label{eq:potential}}
We note that the first and the second term are generated by the magnetic flux  and   the curvature of the internal manifold ($S^2$ in our case), respectively.
While these two terms suffice to stabilize $\phi$, as we will see, the vacuum energy density at $\langle \phi\rangle$, or equivalently, the 4-dimensional cosmological constant $\Lambda_4$,  cannot be positive   with $\Lambda_6=0$.
To obtain the metastable dS vacuum, we need the uplift term induced by the positive $\Lambda_6$ in addition.
% Three terms in \eqref{eq:potential} are natural ingredients of the moduli potential : we have to take them into account regardless of the number of dimensions and the structure of the internal manifold.
% Therefore, the potential $V(\phi)$ given by \eqref{eq:potential} can be used as a simple toy model for the moduli stabilization in string theory \cite{Denef:2007pq}.

For the ease of discussion, we define the following dimensionless quantities,
\dis{v(x)=\frac{4\pi}{M_4^2}V(x),\quad\quad x=\frac{e^{- \frac{\phi}{M_4}}}{R_0^2M_6^2},\quad\quad f^2 = \frac{(g_6N)^2}{M_6^2},\quad\quad \lambda =\frac{\Lambda_6}{M_6^6},\label{eq:Dimless}}
with respect to which \eqref{eq:potential} is rewritten as
\dis{v(x)=\frac{f^2}{32\pi^2}x^3- x^2+\lambda x. \label{eq:DimPot}}
Then one finds that the potential has a local minimum at 
\dis{\langle x\rangle=\frac{e^{- \frac{\langle\phi\rangle}{M_4}}}{R_0^2M_6^2}=\frac{32\pi^2}{3f^2}\Big[1+\sqrt{1-\frac{3f^2}{32\pi^2}\lambda}\Big].\label{eq:<x>}}
Here $M_6^2\langle x\rangle$ is nothing more than $\langle R\rangle^{-2}$, or equivalently, the squared KK mass scale $m_{\rm KK}^2$.
Using   \eqref{eq:Planck4}, the KK mass scale also can be expressed as $m_{\rm KK}=\frac{M_4}{\sqrt{4\pi}}\langle x\rangle$.
This is useful because  it is typical to take $M_4$, instead of $M_6$, to be a fixed value given by $2.4\times 10^{18}$GeV and treat $M_4$ and $\langle R\rangle$  as independent parameters.
\footnote{ Indeed, if we fix the value of $M_6$, $M_4$ becomes $0$ in the limit  $\langle R\rangle \to 0$ hence the EFT assuming  weak gravity breaks down even in the very low energy regime. 
On the other hand, the opposite limit $\langle R\rangle \to \infty$ corresponds to the limit $M_4\to\infty$, indicating the decoupling of gravity in the 4-dimensional EFT description. }
Putting \eqref{eq:<x>} into \eqref{eq:DimPot}, one finds that the value of the potential at $\langle x\rangle$ in the unit of $M_4$ is given by
\dis{v(\langle x\rangle)= \Big(\frac{32\pi^2}{3f^2}\lambda\Big)\Big[1-\frac23 \frac{32\pi^2}{3f^2 \lambda}\Big(1+\Big(1-\frac{3f^2}{32\pi^2}\lambda\Big)^{3/2}\Big)\Big]. \label{eq:vmin}}
We note from \eqref{eq:<x>} that the potential has the local minimum  provided $\frac{3}{32\pi^2}f^2\lambda \leq 1$ : if $\frac{3}{32\pi^2}f^2\lambda$ becomes larger than $1$, the potential   monotonically decreases with respect to $\phi$ (or equivalently,   $1/x$)   without the local minimum.
In other words, the potential in this case shows the runaway behavior.
On the other hand, \eqref{eq:vmin} tells us that $v(\langle x \rangle)$ is positive, i.e.,  the potential is stabilized at the metastable dS vacuum, provided $\frac{3}{32\pi^2}f^2\lambda > \frac34$.
If $\lambda=0$, $v(\langle x \rangle)$ is always negative, then the potential is stabilized at the AdS vacuum.

 %%%%%%%%%%%%%%%%%%%%%%%%%%%%%%%%%%%%%%%%%%%%%%%%%%%%%%%%%%%%%%%%%%%%%%%%%%%%%%%%%%%%%%%%
 \begin{figure}[!t]
 \begin{center}
 \includegraphics[width=0.4\textwidth]{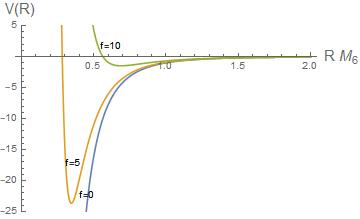}
 \includegraphics[width=0.4\textwidth]{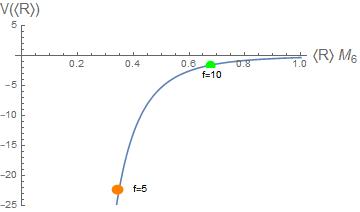}
 \end{center}
\caption{(Left): Behavior of the potential with respect to $R M_6=x^{-1/2}$ when $\lambda=0$. 
 The flux for each plot is given by $f=0, 5$, and $10$, respectively. 
 (Right): Relation between the stabilized values $\langle R\rangle$ and $V(\langle R\rangle)$.
Orange and green dot  indicate the values of $\langle R\rangle$ and $V(\langle R\rangle)$ for $f=5$ and $10$, respectively, which correspond  to plots in the left panel.  
This shows that as $f$  increases,  $\langle R\rangle$ gets larger while $V(\langle R\rangle)$ gets closer to zero. }
\label{Fig:lambda=0}
\end{figure}
%%%%%%%%%%%%%%%%%%%%%%%%%%%%%%%%%%%%%%%%%%%%%%%%%%%%%%%%%%%%%%%%%%%%%%
 
   %%%%%%%%%%%%%%%%%%%%%%%%%%%%%%%%%%%%%%%%%%%%%%%%%%%%%%%%%%%%%%%%%%%%%%%%%%%%%%%%%%%%%%%%
 \begin{figure}[!t]
 \begin{center}
  \includegraphics[width=0.4\textwidth]{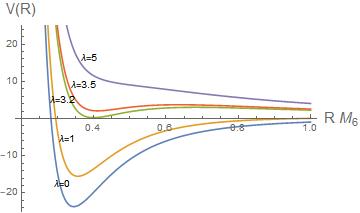}
  \includegraphics[width=0.25\textwidth]{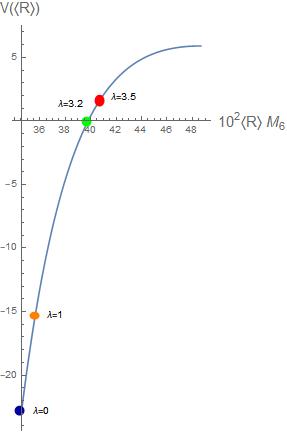}
 \end{center}
\caption{(Left): Behavior of the potential with respect to $R M_6=x^{-1/2}$ when $f=5$. 
  The 6-dimensional cosmological constant in each plot is given by $\lambda=0, 1, 3.2, 3.5$ and $5$, respectively.
 (Right): Relation between   $\langle R\rangle$ and $V(\langle R\rangle)$, showing   that as $\lambda$  increases, both $\langle R\rangle$ and  $V(\langle R\rangle)$ increase.
 In particular, as $\lambda$ becomes larger than $3.2$, $V(\langle R\rangle)$ changes the sign from negative (AdS) to positive (dS). 
Blue, orange, green, and red dot  indicate the values of $\langle R\rangle$ and $V(\langle R\rangle)$ for $\lambda=0, 1, 3.2$ and $3.5$, respectively, which correspond  to plots in the left panel.   }
\label{Fig:lambdadS}
\end{figure}
%%%%%%%%%%%%%%%%%%%%%%%%%%%%%%%%%%%%%%%%%%%%%%%%%%%%%%%%%%%%%%%%%%%%%%

 To investigate the stabilization of the radion in detail, we first consider the case of $\lambda=0$, i.e., the vanishing $\Lambda_6$, where the potential is minimized at $\langle x\rangle=\frac{64\pi^2}{3f^2}$.
Since $\langle x\rangle$ is identified with $1/(M_6\langle R\rangle)^2$, the radius of $S^2$ is given by
\dis{\langle R\rangle=\frac{\sqrt3}{8\pi}\frac{f}{M_6}=\frac{3}{32 \pi^{3/2}}\frac{f^2}{M_4},}
where \eqref{eq:Planck4} is used for the last equality.
Then as mentioned above, the potential at the minimum (in the unit of $M_4$),
 \dis{v(\langle x\rangle)=-\frac{4096\pi^4}{27}\frac{1}{f^4}=-\frac{1}{3 M_6^4 \langle R\rangle^4}=-\frac{4\pi}{3}\frac{1}{M_4^2\langle R\rangle^2},}
 %hence $V(\langle x\rangle)=-\frac{1024 \pi^3}{27}\frac{M_4^4}{f^4}=$,
 is negative, thus the potential is stabilized at the AdS vacuum.
 We note that in the absence of the flux $(f=0)$, $\langle R\rangle$ is given by $0$, at which the potential  becomes negative infinity.
 As evident from plots in Fig. \ref{Fig:lambda=0}, as $f$ increases, $\langle R\rangle$ monotonically increases such that $S^2$ is eventually decompactified, and at the same time, the depth of the potential becomes smaller, approaching the Minkowski vacuum indefinitely.
 Moreover, we infer from \eqref{eq:Planck4} that the stabilized value of the potential $V(\langle x \rangle)=\frac{M_4^4}{4\pi}v(\langle x \rangle)$ is just given by $-\frac{4\pi}{3}M_6^4$.
  If we take $M_6$ to be a fixed value, $V(\langle x \rangle)$ is independent of $\langle R\rangle$.
  However,  it is typical to treat $\langle R\rangle$ and the fixed value $M_4$ as   independent parameters, in which case the qualitative features of $V(\langle x \rangle)$ are the same as those of $v(\langle x \rangle)$ discussed so far.
  As we will see, $M_6$ in this case is interpreted as the `species scale', which gets lower as $\langle R\rangle$ gets larger.

  On the other hand, when $\lambda\ne 0$, the potential is stabilized at the AdS (dS) vacuum for $\frac{3}{32\pi^2}f^2\lambda < \frac34$ ($\frac{3}{32\pi^2}f^2\lambda > \frac34$).
  Suppose  $f$ is fixed but $\lambda$ is allowed to vary.
 Then one finds that the transition between the AdS and the dS vacuum at $\frac{3}{32\pi^2}f^2\lambda = \frac34$ (Minkowski vacuum) is smooth with respect to $\lambda$.
 As can be found in  \eqref{eq:<x>}, for a fixed value of $f$, $\langle R\rangle=1/(M_6 \langle x\rangle^{1/2})$ increases as $\lambda$ increases.
 At the same time, the value of $v(\langle x\rangle)$ given by  \eqref{eq:vmin} increases from the negative to the positive value, thus $|\Lambda_4|$   of the AdS vacuum ($\frac{3}{32\pi^2}f^2\lambda < \frac34$)    gets smaller until the  Minkowski  vacuum ($\frac{3}{32\pi^2}f^2\lambda = \frac34$ hence $|\Lambda_4|=0$) is reached, after which  $|\Lambda_4|$ of the dS vacuum  ($\frac{3}{32\pi^2}f^2\lambda > \frac34$)   gets larger.
 Moreover, the values of $\langle x\rangle$ and $v(\langle x\rangle)$ in the limit  $\lambda \to 0$ coincide with those in the case of $\lambda=0$ we previously discussed.
 Such a behavior of the potential is shown in Fig. \ref{Fig:lambdadS}.
 
\begin{figure}[!t]
 \begin{center}
  \includegraphics[width=0.4\textwidth]{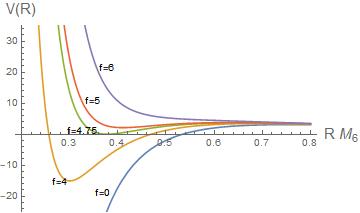}
  \includegraphics[width=0.25\textwidth]{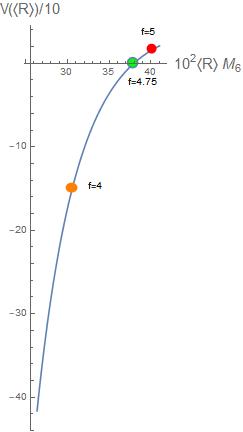}
 \end{center}
\caption{(Left): Behavior of the potential with respect to $R M_6=x^{-1/2}$ when $\lambda=3.5$. 
The flux for each plot is given by  $f=0, 4, 4.75, 5$ and $6$, respectively.
 (Right): Relation between   $\langle R\rangle$ and $V(\langle R\rangle)$, showing   that as $f$  increases, both $\langle R\rangle$  and     $V(\langle R\rangle)$ increase  allowing the transition from the AdS to the dS vacuum. 
  Orange, green, and red dot  indicate the values of $\langle R\rangle$ and $V(\langle R\rangle)$ for $f=4, 4.5$,  and $5$, respectively, which correspond to plots in the left panel.   }
\label{Fig:fdS}
\end{figure}

  We may also fix  $\lambda$ to some finite value, and investigate the behavior of the potential by changing   $f$, as depicted in Fig. \ref{Fig:fdS}. 
  When $f=0$, $\langle x\rangle$ becomes infinity, or equivalently, $\langle R\rangle=0$, at which the value of the  potential is given by negative infinity. 
  This corresponds to the AdS vacuum with the vanishing curvature radius.
  As $f$ increases, the value of the potential increases and when $\frac{3}{32\pi^2}f^2\lambda=\frac34$ is satisfied, we obtain the Minkowski vacuum where the transition from the AdS to the dS vacuum takes place.

    %%%%%%%%%%%%%%%%%%%%%%%%%%%%%%%%%%%%%%%%%%%%%%%%%%%%%%%%%%%%%%%%%%%%%%%%%%%%%%%%%%%%%%%%
%%%%%%%%%%%%%%%%%%%%%%%%%%%%%%%%%%%%%%%%%%%%%%%%%%%%%%%%%%%%%%%%%%%%%%

 Even more interesting case is that both  $f$ and $\lambda$ are allowed to vary but their combination $\frac{3}{32\pi^2}f^2\lambda$ is fixed.
 We postpone the discussion on this case to Sec. \ref{sec:ScaDis} since it is closely relevant to the swampland conjectures concerning the distance conjecture.
 In any case, even if both $f$ and $\lambda$ are nonzero, they cannot be arbitrarily large  since as $\frac{3}{32\pi^2}f^2\lambda$ gets larger than $1$, the potential shows the runaway behavior, i.e.,  monotonically decreases with respect to $R=M_6^{-1}x^{-1/2}$ without the local minimum.
 The values of the stabilized radius and the potential at $\frac{3}{32\pi^2}f^2\lambda=1$ are given by 
  \dis{&\langle x\rangle=\frac{1}{M_6^2\langle R\rangle^2}=\frac{32\pi^2}{3f^2}=\lambda,\quad\quad
v(\langle x\rangle)= \frac{1024 \pi^4}{27} \frac{1}{f^4}=\frac{\lambda^2}{3},}
respectively, which correspond to the maximum values of $\langle R\rangle$ and $v(\langle x\rangle)$ that the metastable dS vacuum can have.

 We also  note that  putting the relation  
 \dis{\frac{3f^2}{32\pi^2}=2(M_6 \langle R\rangle)^2-\lambda(M_6 \langle R\rangle)^4,}
which is obtained from  \eqref{eq:<x>}, into  \eqref{eq:vmin}, we can express $v(\langle x\rangle)$ in terms of $\langle R\rangle$ : 
 \dis{v(\langle x\rangle)=\frac{-1+2 \lambda (M_6 \langle R\rangle)^2}{3(M_6 \langle R\rangle)^4}.}
This is useful in the later discussion.
 
    %%%%%%%%%%%%%%%%%%%%%%%%%%%%%%%%%%%%%%%%%%%%%%%%%%%%%%%%%%%%%%%%%%%%%%%%%%%%%%%%%%%%%%%%
 \begin{figure}[!t]
 \begin{center}
  \includegraphics[width=0.4\textwidth]{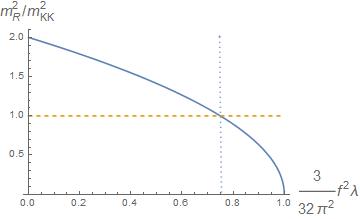}
 \end{center}
\caption{The ratio $m_R^2/m_{\rm KK}^2$ as a function of $\frac{3}{32\pi^2}f^2\lambda$. 
  }
\label{Fig:Ratio}
\end{figure}
%%%%%%%%%%%%%%%%%%%%%%%%%%%%%%%%%%%%%%%%%%%%%%%%%%%%%%%%%%%%%%%%%%%%%%

 Meanwhile, the squared mass of $\phi$, the canonically normalized radion, is given by
 \dis{m_R^2 &=\frac{d^2V}{d\phi^2}\Big|_{\phi=\langle \phi\rangle}=\frac{1}{M_4^2}x\frac{d}{dx}\Big(x\frac{dV}{dx}\Big)\Big|_{x=\langle x\rangle}
 \\
 &=\frac{512\pi^3}{9}\frac{M_4^2}{f^4}\sqrt{1-\frac{3f^2}{32\pi^2}\lambda}\Big(1+\sqrt{1-\frac{3f^2}{32\pi^2}\lambda}\Big)^2,}
 where we use the relation between $\phi$ and $x$ as well as that between $V$ and $v$, both of which can be found in \eqref{eq:Dimless}.
 If $\lambda=0$, it is given by $m_R^2=\frac{2048 \pi^3}{9f^4}M_4^2$, which also can be written as $\frac{128\pi^2}{3f^2}M_6^2$.
 When $\frac{3}{32\pi^2}f^2\lambda=\frac34$ is satisfied, i.e., the radion is stabilized at the Minkowski vacuum, the radion mass is given by $m_R^2=\frac{64\pi^3}{f^4}M_4^2=\frac{16\pi^2}{f^2}M_6^2$, or equivalently, $m_R^2=\frac{\lambda^2}{\pi}M_4^2=2\lambda M_6^2$.
 Of course, the radion becomes massless when  $\frac{3}{32\pi^2}f^2\lambda = 1$, as for $\frac{3}{32\pi^2}f^2\lambda > 1$ the potential is not stabilized but exhibits the runaway behavior.
 It is instructive to compare this with the KK mass scale,   $m_{\rm KK}^2=\langle R\rangle^{-2}=M_6^2\langle x\rangle$,  where $\langle x\rangle$ is given by \eqref{eq:<x>}.
 The   ratio
 \dis{\frac{m_R^2}{m_{\rm KK}^2}=2\sqrt{1-\frac{3f^2}{32\pi^2}\lambda},}
  monotonically decreases from $2$ to $0$ as the combination $\frac{3}{32\pi^2}f^2\lambda$ varies from $0$ to $1$, which is evident from Fig. \ref{Fig:Ratio}.
 In particular, the ratio becomes $1$ at the Minkowski vacuum, where $\frac{3}{32\pi^2}f^2\lambda=\frac34$ is satisfied.
 Therefore, $m_{R}^2$ is always larger than  but not exceeding  twice of  $m_{\rm KK}^2$ in the AdS vacuum while is smaller than $m_{\rm KK}^2$ in the dS vacuum.
 When either $\lambda$ or $f$ is zero, the ratio is fixed to $2$.

 This shows that in the AdS vacuum,  the 4-dimensional EFT description for the radion stabilization is more or less valid since $m_R$ is not so much enhanced compared to $m_{\rm KK}$ hence only a few  KK modes are found at the energy scale $m_R$.
 Moreover, the ratio $m_R^2/m_{\rm KK}^2$ does not diverge, implying that even if $m_{\rm KK}$ gets closer to $0$, it does not decrease more rapidly compared to $m_R$, so the radion stabilization may  not be obstructed by the descent of the KK modes.
However, even in this case, the UV cutoff for any other 4-dimensional EFT description can decrease rapidly in the limit $m_{\rm KK}\to 0$.
Indeed, more realistic model can contain the dynamics which is not directly relevant to the radion stabilization, but well described in the 4-dimensional EFT framework.
  For example, the size of $\Lambda_4$ resulting from the moduli stabilization and that of the electroweak scale can be connected only in the indirect way.
  For this reason, we will regard the vanishing limit of $m_{\rm KK}$ as a signal that the generic 4-dimensional EFT description becomes invalid, as well known.

\section{Distance conjecture : scale separation and scaling law}
\label{sec:ScaDis}

 Since two extra dimensions are compactified on $S^2$, the 6-dimensional Einstein-Maxwell theory we reviewed in Sec. \ref{sec:6DEM}  contains the KK tower as a natural tower of states.
 When the KK mass scale $m_{\rm KK}$ becomes zero, the KK tower descends from UV, invalidating  the 4-dimensional EFT description.
 Recent distance conjecture claims that such a descent of a tower of states is what happens when the EFT is close to the corner of the landscape.
 Moreover, it is also  conjectured that a tower of states invalidating the EFT descends from UV as the size of $\Lambda_4$ gets smaller, emphasizing that the (A)dS vacuum in the vanishing limit of $\Lambda_4$ cannot be continuously interpolated into the Minkowski vacuum in the EFT framework    \cite{Lust:2019zwm}.
 In this section, we discuss the conjectures concerning a tower of states  in the context of the 6-dimensional Einstein-Maxwell theory, in which both  $m_{\rm KK}$ and $\Lambda_4$ are determined by the radion stabilization thus the flux and the uplift.

 We begin our discussion with the scale separation between   the IR mass scale  obtained from  $\Lambda_4$ and $m_{\rm KK}$.
 When the background geometry is given by dS space, the natural IR mass scale is the Hubble parameter $H$ satisfying $\Lambda_4=3M_4^2H^2$, which indeed is the inverse of the horizon radius.
 In the case of the AdS background, the inverse of the curvature radius, which will be also denoted by $H$, is the natural IR mass scale.
 This can be justified by considering the metric of AdS space \cite{Susskind:1998dq} (see also  \cite{Zwiebach:2004tj, Calderon-Infante:2023ler})  
 \dis{ds^2=\frac{1}{H^2}\Big[-\Big(\frac{1+r^2}{1-r^2}\Big)dt^2+\frac{4}{(1-r^2)^2}(dr^2+r^2d\Omega^2)\Big],\label{eq:AdSmet}}
 where $r \in [0,1)$ and $r=1$ corresponds to the boundary.
 When we imagine the surface of constant $r$, the area of the surface and the volume of the region surrounded by the surface are given by
 \dis{&A=4\pi\Big(\frac{2 r}{H(1-r^2)}\Big)^2
 \\
 &V=4\pi \int dr r^2\Big(\frac{2 }{H(1-r^2)}\Big)^3 = \frac{4\pi}{2H^3}\Big[\frac{2r(1+r^2)}{(1-r^2)^2}+\log\Big(\frac{1-r}{1+r}\Big)\Big],\label{eq:AdSgeom}}
 respectively.
  Then the ratio $A/V$  is a monotonically decreasing function of $r$ and minimized at the boundary ($r=1$) where $A/V= 2H$.
 
     %%%%%%%%%%%%%%%%%%%%%%%%%%%%%%%%%%%%%%%%%%%%%%%%%%%%%%%%%%%%%%%%%%%%%%%%%%%%%%%%%%%%%%%%
 \begin{figure}[!t]
 \begin{center}
  \includegraphics[width=0.4\textwidth]{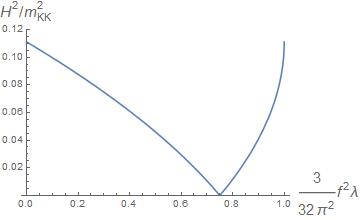}
 \end{center}
\caption{The ratio $H^2/m_{\rm KK}^2$ as a function of $\frac{3f^2\lambda}{32\pi^2}$. 
  }
\label{Fig:HKKRatio}
\end{figure}
%%%%%%%%%%%%%%%%%%%%%%%%%%%%%%%%%%%%%%%%%%%%%%%%%%%%%%%%%%%%%%%%%%%%%%

In terms of $H$, the value  of $\Lambda_4$ in dS space and that in AdS  space are written as $3H^2 M_4^2$ and $-3H^2 M_4^2$, respectively : as (A)dS space gets closer to Minkowski space,    $|\Lambda_4|$ gets smaller, and $H$ becomes $0$.
 In the 6-dimensional Einstein-Maxwell theory, the value of $H$  is given by
 \dis{H^2&=\frac{|V(\langle\phi\rangle)|}{3 M_4^2}=\frac{M_4^2}{12\pi}|v(\langle x\rangle)|
 \\
 &=\frac{M_4^2}{12\pi}\Big(\frac{32\pi^2}{3f^2}\lambda\Big)\Big|1-\frac23 \frac{32\pi^2}{3f^2 \lambda}\Big(1+\Big(1-\frac{3f^2}{32\pi^2}\lambda\Big)^{3/2}\Big)\Big|. \label{eq:H2}}
 We note that the potential for $f=0$ (the vanishing magnetic flux) is minimized at $\langle R\rangle=0$ where $V=-\infty$ hence $H=\infty$.
 As mentioned in Sec. \ref{sec:6DEM}, this will be interpreted as the AdS vacuum in the limit of the vanishing curvature radius.
 Using the expression for $\langle x\rangle=m_{\rm KK}^2/M_6^2$ given by \eqref{eq:<x>}  and the fact that $M_4^2=4\pi M_6^4/m_{\rm KK}^2$ (see \eqref{eq:Planck4}), we obtain the ratio
 \dis{\frac{H^2}{m_{\rm KK}^2}&=\frac13 \Big(\frac{3f^2}{32\pi^2}\lambda\Big) \frac{\Big|1-\frac23 \frac{32\pi^2}{3f^2 \lambda}\Big(1+\Big(1-\frac{3f^2}{32\pi^2}\lambda\Big)^{3/2}\Big)\Big|}{\Big(1+\sqrt{1-\frac{3f^2}{32\pi^2}\lambda}\Big)^2 }
 \\
 &=\frac19|-1+2\lambda(M_6\langle R\rangle)^2|
 \\
 &=\frac19\Big|-1+ \frac{\lambda}{\pi}\frac{M_4}{m_{\rm KK}} \Big|,\label{eq:HKKratio}}
 the behavior of which as a function of $\frac{3f^2\lambda}{32\pi^2}$ is depicted in Fig. \ref{Fig:HKKRatio}.
 From this, one finds that when $\frac{3f^2\lambda}{32\pi^2} =1$, i.e., the dS background is about to be destabilized by the runaway behavior of the potential, $H^2/m_{\rm KK}^2=1/9$.
  The same ratio $H^2/m_{\rm KK}^2=1/9$ is also obtained in the limit  $\frac{3f^2\lambda}{32\pi^2} \to 0$.
  In these cases, The IR mass scale  $H$ is of the same order as $m_{\rm KK}$, so the scale separation does not occur.
  
 In particular, when $\lambda=0$, the value of $H^2/m_{\rm KK}^2$ is $1/9$ for any finite value of   $f$ since $\frac{3f^2\lambda}{32\pi^2}=0$ is satisfied, while the radion is always stabilized at the AdS vacuum.
 Then we obtain the scaling law obeyed by  $\Lambda_4$ and $m_{\rm KK}$ given by $m_{\rm KK}=|3\Lambda_4|^{1/2}/M_4$.
 The exponent $1/2$ here is claimed to be the upper bound on the exponent in the AdS/dS distance conjecture \cite{Lust:2019zwm}, which can be found in the supersymmetric AdS vacuum.
 The scaling law shows that as $|\Lambda_4| \to 0$, i.e., the AdS background gets closer to  Minkowski space, $m_{\rm KK}$ becomes zero, invalidating the 4-dimensional EFT description.
 In terms of $\langle \phi\rangle$, the relation $m_{\rm KK}=\frac{M_4}{\sqrt{4\pi}}\langle x\rangle$ with $x\sim e^{-\phi/M_4}$ indicates that $|\Lambda_4|/M_4^4 \sim  e^{-2\phi/M_4}$.
We note that since the size of both  $\langle x\rangle  \sim f^{-2}$ and $ v(\langle x\rangle)\sim f^{-4}$ are controlled by the flux $f$, the smallness of $|\Lambda_4|$ indicating the low KK mass scale can be understood as a result of the large flux, as also discussed in Sec. \ref{sec:6DEM}.
%Indeed, while the flux term in $v(x)$ given by $\frac{f^2}{32\pi^2}x^3$ is proportional to $f^2$, its value at the stabilized minimum is proportional to $f^{-4}$, which is of the same order as the size of $v(\langle x\rangle)$.
%One may express this as the relation $\frac{f^2}{32\pi^2}\langle x\rangle^3=\frac23 \langle x\rangle^2$.

 On the contrary,  when $\lambda \ne 0$, as can be inferred from the last expression in \eqref{eq:HKKratio}, the ratio between $H$ and $m_{\rm KK}$ is no longer a constant  unless $\lambda$ is close to $0$.
Indeed, as discussed in Sec. \ref{sec:6DEM} and summarized in   Fig. \ref{Fig:lambdadS} and Fig. \ref{Fig:fdS}, given the fixed value of $f$ ($\lambda$),  the transition between the AdS vacuum and the dS vacuum through the Minkowski vacuum  is smooth with respect to $\lambda$ ($f$).
The KK mass scale does not vanish at the Minkowski vacuum ($\frac{3f^2\lambda}{32\pi^2} = \frac34$) in this case, and as is evident from Fig. \ref{Fig:HKKRatio}, $H$ (hence $|\Lambda_4|$) decreases even more rapidly compared to $m_{\rm KK}$, thus  the KK tower can be decoupled from the  EFT at the IR mass scale.
This is quite different from the prediction of  the  AdS/dS distance conjecture and seems to realize the scenario similar to the KKLT \cite{Kachru:2003aw} or the large volume scenario \cite{Balasubramanian:2005zx}, in which the radion is mainly stabilized  by the flux and then slightly uplifted, while the KK tower does not descend from UV even in the Minkowski limit.

The situation can be changed if we allow both $f$ and $\lambda$ to vary but fix their combination $\frac{3f^2 \lambda}{32\pi^2}$  to some value smaller than  $1$ (to avoid the runaway potential).
This is equivalent to fixing the combination $\lambda (M_6\langle R\rangle)^2=(\lambda/\sqrt{4\pi})M_4 \langle R\rangle=\lambda \langle x\rangle^{-1}$, which is evident from \eqref{eq:HKKratio} or  the fact that $\lambda \langle x\rangle^{-1}$ depends on $f$ and $\lambda$ through the combination $\frac{3f^2 \lambda}{32\pi^2}$  only (see \eqref{eq:<x>}).
In this case,  $H^2/m_{\rm KK}^2$ becomes a constant   smaller than $1/9$, resulting in the scaling law $m_{\rm KK} \sim |\Lambda_4|^{1/2}$ as in the case of  $\lambda=0$ or $1$.
Of course, depending on the value of $\frac{3f^2 \lambda}{32\pi^2}$ , the vacuum we consider can be not only AdS or dS, but also Minkowski.
Moreover, under the fixed value of $\frac{3f^2 \lambda}{32\pi^2}$, $f^2$ gets larger  if we choose smaller $\lambda$.
Equivalently,  the value of $\lambda (M_6\langle R\rangle)^2=\frac{M_4}{\sqrt{4\pi}}\lambda  \langle R\rangle$ can be fixed if we take  $\langle R\rangle$ to be infinitely large (hence $m_{\rm KK}\to 0$) in the limit $\lambda \to 0$.

     %%%%%%%%%%%%%%%%%%%%%%%%%%%%%%%%%%%%%%%%%%%%%%%%%%%%%%%%%%%%%%%%%%%%%%%%%%%%%%%%%%%%%%%%
 \begin{figure}[!t]
 \begin{center}
  \includegraphics[width=0.4\textwidth]{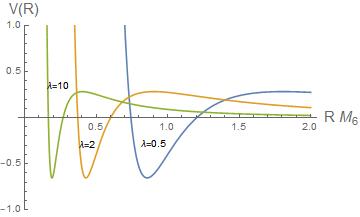}
 \end{center}
\caption{Radion potentials giving the same value of the AdS cosmological constant under the fixed value of $M_6$. 
While $\frac{3f^2 \lambda}{32\pi^2}$ is fixed to $0.2$, the value of $\lambda$ for each plot is chosen to be $0.5$, $2$ and $10$, respectively.
  }
\label{Fig:constcomb}
\end{figure}
%%%%%%%%%%%%%%%%%%%%%%%%%%%%%%%%%%%%%%%%%%%%%%%%%%%%%%%%%%%%%%%%%%%%%%

 Now, from \eqref{eq:vmin},  the stabilized value of the potential which is identified with $\Lambda_4$ can be  written as
  \dis{\Lambda_4&=V(\langle x\rangle) =\frac{M_4^4}{4\pi}v(\langle x\rangle)
  \\
  &=\frac{M_4^4}{4\pi}\lambda^2 \Big(\frac{32\pi^2}{3f^2\lambda}\Big)\Big[1-\frac23 \frac{32\pi^2}{3f^2 \lambda}\Big(1+\Big(1-\frac{3f^2}{32\pi^2}\lambda\Big)^{3/2}\Big)\Big]
  \\
  &= 4\pi M_6^8  (\lambda\langle R\rangle^2)^2 \Big(\frac{32\pi^2}{3f^2\lambda}\Big)\Big[1-\frac23 \frac{32\pi^2}{3f^2 \lambda}\Big(1+\Big(1-\frac{3f^2}{32\pi^2}\lambda\Big)^{3/2}\Big)\Big]. \label{eq:cosconst}}
   If we take $M_6$ to be a fixed value,  $\lambda \langle R\rangle^2$ in the last expression of \eqref{eq:cosconst}, hence $\Lambda_4$ is  completely determined by the value of $\frac{3f^2 \lambda}{32\pi^2}$. 
    We can compare cases of different values of $\lambda$ giving the same value of $\frac{3f^2 \lambda}{32\pi^2}$ (hence the same value of  $\Lambda_4$) as follows (see also Fig. \ref{Fig:constcomb}).
  When we separate the radion stabilization into two steps,  the stabilization of the radion at the AdS vacuum by $f$ and the uplift by $\lambda$, the larger value of $f$ stabilizes the radion  at the AdS vacuum of the   smaller depth, but since $\lambda$ becomes smaller to give the same value of $\frac{3f^2 \lambda}{32\pi^2}$, the potential is uplifted by the  smaller amount, giving the same value of $\Lambda_4$.
  \footnote{The final value of $\Lambda_4$ is not necessarily positive : the sign of $\Lambda_4$ is determined by the value of $\frac{3f^2 \lambda}{32\pi^2}$.}
  In particular,   we can take the limit $\lambda\to 0 $ without changing  $\Lambda_4$, in which  both $f$ and $\langle R\rangle$ becomes infinity, resulting in $m_{\rm KK}\to 0$.
  Since the KK  tower descends from UV in this case, the 4-dimensional EFT description   is not reliable.
 We can find the similar situation in more realistic model like the KKLT or the large volume scenario, in which the uplift potential is generated by $\overline{\rm D3}$-branes at the tip of the Klebanov-Strassler throat \cite{Blumenhagen:2022zzw, Seo:2023ssl}.
 In this case, the uplift potential and the tower mass scale $m$ satisfy the scaling law $m \sim V_{\rm up}^\alpha$. 
 The value of $\alpha$ is given by either $1/3$ or $1/4$, depending on the type of tower (the KK or the string tower) and how strongly the throat is warped.

     %%%%%%%%%%%%%%%%%%%%%%%%%%%%%%%%%%%%%%%%%%%%%%%%%%%%%%%%%%%%%%%%%%%%%%%%%%%%%%%%%%%%%%%%
 \begin{figure}[!t]
 \begin{center}
  \includegraphics[width=0.4\textwidth]{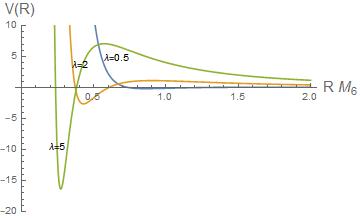}
 \end{center}
\caption{Radion potentials for different choices of $\lambda$ under the fixed value of  $M_4$. 
While $\frac{3f^2 \lambda}{32\pi^2}$ is fixed to $0.2$, the value $\lambda$ for each plot is chosen to be $0.5$, $2$ and $5$, respectively.
  }
\label{Fig:constcomb2}
\end{figure}
%%%%%%%%%%%%%%%%%%%%%%%%%%%%%%%%%%%%%%%%%%%%%%%%%%%%%%%%%%%%%%%%%%%%%%

  On the other hand, if we fix the value of $M_4$,  the second expression of \eqref{eq:cosconst} tells us that given the fixed value of $\frac{3f^2 \lambda}{32\pi^2}$, the smaller value of $\lambda$ leads to the smaller size of $\Lambda_4$ (Fig. \ref{Fig:constcomb2}).
   But since the sign of $\Lambda_4$ cannot be changed by varying $\lambda$ in this case,    the transition between the  AdS and the dS vacuum does not take place.
  Moreover, while the limit $\lambda\to 0$  corresponds to the Minkowski limit $\Lambda_4 \to0$ of (A)dS space (the sign of $\Lambda_4$ depends on the value of $\frac{3f^2 \lambda}{32\pi^2}$), $\langle R\rangle$ becomes infinity as well, thus $m_{\rm KK} \to 0$.
  More concretely,  from \eqref{eq:<x>}, the KK mass scale can be written as
  \dis{m_{\rm KK}=\frac{M_4}{\sqrt{4\pi}}\langle x\rangle=\frac{M_4}{\sqrt{4\pi}}\lambda \Big(\frac{32\pi^2}{3f^2 \lambda}\Big)\Big[1+\sqrt{1-\frac{3f^2}{32\pi^2}\lambda}\Big], }  
 so when $M_4$ and   $\frac{3f^2 \lambda}{32\pi^2}$ are fixed, $m_{\rm KK}$ becomes $0$ as $\lambda \to 0$.
 Therefore, when $\lambda$ is very close to $0$, the 4-dimensional EFT   is no longer reliable as  the KK tower descends from UV, which is consistent with the claim  of  the AdS/dS distance conjecture.
 Meanwhile, when $\frac{3f^2 \lambda}{32\pi^2}$ is fixed to $\frac34$, the radion is stabilized at the Minkowski vacuum ($\Lambda=0$), regardless of the size of $\lambda$ : the transition between the Minkowski and the (A)dS vacuum  does not take place.
 Even in this case,     the radion stabilization in the limit of $\lambda \to 0$ is not reliable since  $m_{\rm KK}=\frac{M_4}{\sqrt{\pi}}\lambda$ becomes $0$ as well.
   In terms of the moduli stabilization, this indicates that the realization of the Minkowski vacuum through the tiny uplift from very shallow AdS potential generated by the large flux  is invalidated by the descent of the KK tower.

 We close this section with the remark on the species scale $\Lambda_{\rm sp}$, above which  gravity is no longer weakly coupled : denoting  the number of states below $\Lambda_{\rm sp}$ by $N_{\rm sp}$, $\Lambda_{\rm sp}$ is defined by the requirement that the gravity coupling $N_{\rm sp}\Lambda_{\rm sp}/M_4^2$ becomes ${\cal O}(1)$, i.e., $\Lambda_{\rm sp}=M_4/\sqrt{N_{\rm sp}}$ \cite{Dvali:2007hz, Dvali:2007wp}.
 Then $\Lambda_{\rm sp}$ is a natural cutoff scale of the EFT in which the gravitational interaction is weak.
  When $m_{\rm KK}$ decreases,  the descent of the KK tower leads to the increase in $N_{\rm sp}$, thus the decrease in  $\Lambda_{\rm sp}$.
  In our case, two extra dimensions are compactified on $S^2$, then the squared mass of the KK mode  is written as $m_{\mathbf n}^2=\langle R\rangle^{-2} (n_1^2 +n_2^2)=m_{\rm KK}^2 (n_1^2 +n_2^2)$, 
with $n_1$ and $n_2$ integers (for more generic case in which the sizes of two extra dimensions are different, see, e.g., \cite{Castellano:2021mmx} or Appendix A of  \cite{Seo:2023xsb}).  
Since we expect that $m_{\rm KK} \ll \Lambda_{\rm sp}$, $N_{\rm sp}$ is approximated by a quarter  of the area of the disc with radius $\Lambda_{\rm sp}/m_{\rm KK}$.
Omitting ${\cal O}(1)$ coefficient,   $N_{\rm sp}$ is estimated as $ \Lambda_{\rm sp}^2/m_{\rm KK}^2$.
Combining this with the relation  $\Lambda_{\rm sp}=M_4/\sqrt{N_{\rm sp}}$, we obtain
\dis{N_{\rm sp}=\frac{M_4}{m_{\rm KK}},\quad\quad \Lambda_{\rm sp}=m_{\rm KK}^{1/2} M_4^{1/2}.} 
Moreover, using the relation $M_4=\sqrt{4\pi}M_6^2/m_{\rm KK}$ (see \eqref{eq:Planck4}), one finds that $\Lambda_{\rm sp}=M_6$.  
That is,  the strong gravitational interaction becomes evident above the more fundamental gravity scale $M_6$ (see also \cite{Castellano:2022bvr} for more discussion).
This indeed is consistent with the well known proposal that the very high mass scale $M_4$ may be the result of the large   extra dimensions, in which case the Planck mass of the higher dimensional UV completion ($M_6$ in our case)  is low  \cite{Arkani-Hamed:1998jmv}.
It is also remarkable that $N_{\rm sp}$ is nothing more than $\langle x\rangle^{-1}$, since
\dis{N_{\rm sp}\sim M_4\langle R\rangle=\sqrt{4\pi}(M_6\langle R\rangle)^2=\frac{\sqrt{4\pi}}{\langle x\rangle}.}
Therefore, using \eqref{eq:<x>} we can parametrize $N_{\rm sp}$ in terms of $f$ and $\lambda$.
For example, when $\lambda=0$, the relation $N_{\rm sp}\sim \langle x\rangle^{-1}=\frac{3f^2}{64\pi^2}$ is satisfied, showing that the number of KK modes below $\Lambda_{\rm sp}=M_6$ increases as $f$ increases.
Meanwhile, when $\lambda\ne 0$ and the value of $\frac{3 f^2\lambda}{32\pi^2}$ is fixed, the combination $\lambda (M_6\langle R\rangle)^2=\lambda \langle x\rangle^{-1}$ is also fixed, which leads to $N_{\rm sp}\sim 1/\lambda$ up to ${\cal O}(1)$ constant.
This explicitly shows the descent of the KK tower   in the limit  $\lambda \to 0$.

 \section{Vacuum transition and the bubble of nothing   }
 \label{sec:BoN}
 
As we have seen in Sec. \ref{sec:6DEM}, the 6-dimensional Einstein-Maxwell theory contains essential ingredients of the moduli stabilization,  the flux and the uplift  parametrized by $f$ and $\lambda$, respectively.
  A large number of (meta)stable vacua in the landscape can be realized by various choices of  $f$ and $\lambda$, provided the EFT description is protected from the descent of the KK tower.
 Moreover, the 6-dimensional Einstein-Maxwell theory admits the instanton solution describing the nucleation of the bubble, which  allows   a transition between vacua having different  values of $\Lambda_4$   \cite{Blanco-Pillado:2009lan, Brown:2010bc, Brown:2010mf, Brown:2011gt}.
 Among possible solutions for the bubble,   the `bubble of nothing'  is of particular interest in light of the cobordism conjecture \cite{McNamara:2019rup}.
 In this solution,  noncompact spacetime cannot extend inside the bubble as it pinches off at the bubble wall, where the radius of the extra dimensions shrinks to zero size   \cite{Witten:1981gj}.

 To investigate various aspects of the bubble solution including the bubble of nothing in detail, we first note that in the bubble solution of the 6-dimensional Einstein-Maxwell theory, the magnetic flux through $S^2$ is `discharged' by the black 2-brane at the bubble wall.
  Here the black 2-brane is extended over  noncompact $(1+2)$-dimensions and described by the metric \cite{Blanco-Pillado:2009lan}
 \dis{ds^2=\Big(1-\frac{\sqrt3 g_6}{8\pi M_6^2 \rho}\Big)^{2/3}(-dt^2+dx^2+dy^2)+\Big(1-\frac{\sqrt3 g_6}{8\pi M_6^2 \rho}\Big)^{-2} d\rho^2+\rho^2(d\psi^2+\sin^2\psi d\omega^2),\label{eq:brane}}
 which is the solution to the Einstein equation in the presence of the magnetic flux through $S^2$, i.e., the electromagnetic field strength given by $F_2=\frac{g_6}{4\pi}\sin \psi d\psi\wedge d\omega$.
 Then   the flux quanta outside and inside the bubble   are given by  $N$ (see \eqref{eq:EMsol}) and $N-n$, respectively, where $n$ is the number of branes. 
  Even in this case, the values of the 6-dimensional cosmological constant $\Lambda_6$ (hence   $\lambda$) inside   and   outside the bubble   are the same, since $\Lambda_6$ is not changed by the black 2-brane. 
 As a result, the values of $\Lambda_4$ inside and outside the bubble  are different and the bubble solution well describes the transition between two vacua  through the tunnelling \cite{Coleman:1980aw}. 
 We also expect that another bubble can be nucleated inside  the bubble, which leads to the sequential change in  $\Lambda_4$ along some particular direction (normal to the bubble wall) in the coordinate space.

  When the flux inside the bubble vanishes ($f=0$),  $\Lambda_4$ becomes negative infinity, hence the geometry   inside  the bubble corresponds to AdS space with the vanishing curvature radius ($H=\infty$).
  In this case, $\langle R\rangle=0$  and as can be inferred from   \eqref{eq:AdSgeom},  the volume vanishes for any finite value of $r$.
 Then the region inside the bubble  may be regarded as being empty, i.e., not filled with spacetime.
  From this, \cite{Brown:2011gt} claimed that the bubble of vanishing flux can be identified with the bubble of nothing, which was extensively studied in  \cite{Brown:2010bc, Brown:2010mf, Brown:2011gt}.

 Before proceeding, we note here about the reliability of the claim in \cite{Brown:2011gt}.
 Whereas the bubble is the classical solution, the point of vanishing $\langle R\rangle$ corresponds to the spacetime singularity, where the length scale provided by the curvature becomes much smaller than the cutoff length scale, say, the Planck length or the string length scale, hence the EFT based on the seimclassical approximation breaks down.
 Another way to see the invalidity of the EFT is to notice that if the membrane wrapping $S^2$ exists, its tension becomes extremely small as $\langle R\rangle$ gets close to zero.
 This means the descent of a tower of states which was not present in  the EFT  from UV.
 On the other hand, the recent cobordism conjecture  \cite{McNamara:2019rup} argues that any internal manifold in the EFT consistent with quantum gravity can shrink to the nothing without the topological obstruction.
 This suggests that if we assume our 6-dimensional Einstein-Maxwell theory to be consistent with quantum gravity, the point of vanishing $\langle R\rangle$ can be still interpreted as the `end of the world' beyond which spacetime no longer extends.
 At this point, the value of $\langle R\rangle$ may not actually vanish, but instead the defect of the small but finite size can be located, as considered in \cite{Friedrich:2023tid}.
 From this observation, we regard the vanishing limit of $f$ as the end of the world in the semiclassical approximation, beyond which is just the nothing.

 If $\Lambda_6=0$ ($\lambda=0$), as we discussed in Sec. \ref{sec:6DEM} (and summarized in Fig. \ref{Fig:lambda=0}), the radion is always stabilized at the AdS vacuum where $\Lambda_4$ is  negative.
   The decrease in the flux by the black 2-brane leads to the nucleation of the bubble in  which the geometry is given by AdS space with the smaller value of $\Lambda_4$ (or equivalently, the larger value of $H$).
Then through the sequential nucleation of the bubbles lowering $\Lambda_4$, the AdS vacuum with $\Lambda_4=-\infty$ can be reached, which is interpreted as a nucleation of the bubble of nothing.
Transition to the bubble of nothing from dS or Minkowski space can be realized when $\Lambda_6 \ne 0$ ($\lambda \ne 0$ : see also discussion in Sec. \ref{sec:6DEM} which is summarized in Fig. \ref{Fig:lambdadS}).
Since the nucleation of the bubble just changes $f$, but not $\lambda$,  the transitions between dS, Minkowski, and AdS space are not obstructed by the descent of the KK tower.

 These features are evident in the thin wall approximation, in which the geometry inside (or outside) the bubble   is well described by (Euclidean) (A)dS space, except for the narrow region near the bubble wall.
 Moreover, we assume that the 4-dimensional part of the metric \eqref{eq:EMsol} describing the geometry inside the bubble  respects the spatial isotropy thus can be written as
 \dis{g_{\mu\nu}dx^\mu dx^\nu=d\xi^2+a(\xi)^2d\Omega_3^2,} 
where we use the Euclidean signature since the bubble is an instanton solution.
In addition, $\phi$ depends only on $\xi$.
We also expect that the metric describing the region outside the bubble   can be written in  the same form, but  $\Lambda_4$ becomes different as the flux outside and inside the bubble are different.
Restricting our attention to the region inside the bubble, the action is given by
\dis{S_E^{\rm in}&=\int d^4 x\sqrt{g}\Big[-\frac{M_4^2}{2}{\cal R}+\frac12 g^{ij}\partial_i\phi \partial_j \phi +V(\phi)\Big]
\\
&=2\pi^2 \int d\xi \Big[3M_4^2(a^2a''+a{a'}^2-a)+a^3\Big(\frac12 {\phi'}^2+V(\phi)\Big)\Big],\label{eq:action1}}
where the prime indicates the derivative with respect to $\xi$.
 Thus, $a(\xi)$ and $\phi(\xi)$ satisfy
 \dis{&\frac{3({a'(\xi)}^2-1)}{a(\xi)^2}=\frac{1}{M_4^2}\Big(\frac{\phi'(\xi)^2}{2}-V(\phi)\Big)
 \\
 &\phi(\xi)''+3\frac{a'(\xi)}{a(\xi)}\phi'(\xi)=\frac{dV(\phi)}{d\phi}.\label{eq:EoM}} 
In the thin wall approximation, the geometry inside the bubble   is more or less well described by the solution $\phi=\langle R\rangle=$constant (hence  $V(\phi)=V(\langle R\rangle)$), i.e., the stabilized radion.
 Depending on the sign of $V(\langle R\rangle)$, $a(\xi)$ describes the Euclidean  version of AdS ($V(\langle R\rangle)<0)$, Minkowski ($V(\langle R\rangle)=0$), and dS ($V(\langle R\rangle)>0$) space, respectively.
 Since $|V(\langle R\rangle)|=3M_4^2 H^2$, $a(\xi)$ in the AdS vacuum ($V(\langle R\rangle<0$) is given by
 \dis{a(\xi)=\frac{1}{H}\sinh[H(\xi-\xi_0)],\label{eq:aAdS}} 
whereas $a(\xi)$ in the dS vacuum ($V(\langle R\rangle >0$) is given by
 \dis{a(\xi)=\frac{1}{H}\cos[H(\xi-\xi_0)].\label{eq:adS}}
 From \eqref{eq:HKKratio}, $H$ can be written as
\dis{H=\frac{1}{3\langle R\rangle}\Big|\frac{\lambda}{\pi}M_4 \langle R\rangle-1\Big|^{1/2}} 
  in both cases. 
  For   Euclidean Minkowski space ($V(\langle R\rangle)=0$), we just have $a(\xi)=\xi$.
  
  The action for the whole region, which includes the bubble wall  as well as the regions inside and outside the bubble, gives the profiles of $a(\xi)$ and $\phi(\xi)$ describing the interpolation between two different vacua.
  Even though both $a(\xi)$ and $\phi(\xi)$ are continuous over the whole range of $\xi$, their derivatives  $a'(\xi)$ and $\phi'(\xi)$ show  discontinuity at the bubble wall in the thin wall approximation.
  This is well explored in \cite{Brown:2010mf}, which is also reviewed in App. \ref{app:wholesol}  to make the discussion self-contained.
  Moreover, whereas we are mainly interested in so-called  the `down tunnelling', the nucleation of the bubble with the smaller value of $\Lambda_4$, there can be the `up tunnelling', the nucleation of the bubble with the larger value of $\Lambda_4$.
  Regarding this, we note that   Euclidean dS space is compact, leading to the finite value of the Euclidean action $S_E$ even if the integration is taken over the region outside the bubble.
   In contrast, Euclidean AdS or Minkowski space is noncompact, hence  $S_E$ for the region outside the bubble   diverges. 
  As pointed out in \cite{Brown:2011gt} (and also \cite{Coleman:1980aw}), such a difference prevents the transition from Minkowski or AdS space to the vacuum with larger $\Lambda_4$.
  This is also reviewed in App.  \ref{app:wholesol}.

     %%%%%%%%%%%%%%%%%%%%%%%%%%%%%%%%%%%%%%%%%%%%%%%%%%%%%%%%%%%%%%%%%%%%%%%%%%%%%%%%%%%%%%%%
 \begin{figure}[!t]
 \begin{center}
  \includegraphics[width=0.4\textwidth]{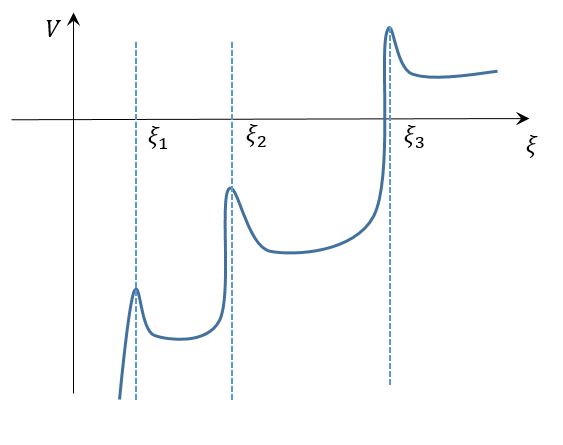}
 \end{center}
\caption{The qualitative profile of $V(\phi(\xi))$ in the sequential nucleation of bubbles.
The flux changing black 2-branes are located at $\xi_1$, $\xi_2$, and $\xi_3$, respectively, and the value of $\Lambda_4$ in each bubble, i.e., the region  between branes, is almost constant in the thin wall approximation.
  }
\label{Fig:potential}
\end{figure}
%%%%%%%%%%%%%%%%%%%%%%%%%%%%%%%%%%%%%%%%%%%%%%%%%%%%%%%%%%%%%%%%%%%%%%

\begin{figure}[!t]
 \begin{center}
  \includegraphics[width=0.4\textwidth]{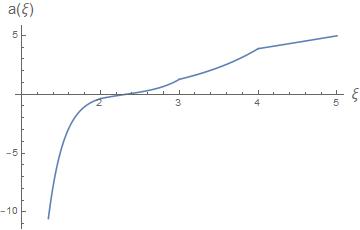}
 \end{center}
\caption{Profile of $a(\xi)$ describing the transition between AdS spaces, where   $H$ increases (hence $\Lambda_4 (<0)$ decreases) discontinuously along the direction of $\xi \to 0$ : the values of $H$ are taken to be $5$ for $1<\xi<2$, $3$ for $2<\xi<3$, $1$ for $3<\xi<4$, and $0.1$ for $4<\xi<5$, respectively, and $\xi_0^4=0.2$.
  }
\label{Fig:AdSprof}
\end{figure}

When the sequential nucleation of the bubbles  takes place,   the change in $\Lambda_4$  in the moduli space (determined by the value of $\langle \phi\rangle$) also can be realized in the real coordinate space along $\xi$ (see Fig. \ref{Fig:potential}).
In particular, consider  the transition toward the nothing, the AdS vacuum with $H=\infty$, along the direction of $\xi \to 0$   through the sequential nucleation of the bubbles.
To describe this, suppose the geometry before the nucleation of the bubble is given by AdS space and the bubble walls are located at $\xi_1, \xi_2, \cdots,$ respectively.
Then $\Lambda_4$ can get smaller through the sequential nucleation of the bubbles until it reaches the negative infinity, i.e., the `nothing'.
\footnote{In fact, the production of the bubble with the larger value of $\Lambda_4$ in AdS space is suppressed.
In particular, As mentioned above, the nucleation of the small bubble with the larger value of $\Lambda_4$ in the noncompact Euclidean AdS space is forbidden, as argued in \cite{Brown:2011gt}  and also reviewed in App. \ref{app:wholesol}.}  
Denoting the value of $H$ in the region $\xi \in (\xi_i, \xi_{i+1})$ by $H_i$, one can infer from \eqref{eq:aAdS} that $a(\xi)$ in this  region is written as
\dis{a_i(\xi)=\frac{1}{H_i}\sinh[H_i(\xi-\xi^{i}_0)].}
We note that in the bubble of nothing, $|a(\xi)|$  diverges since $H\to \infty$.
Ignoring the profile of $a(\xi)$ at the bubble wall which is assumed to be very narrow, it is a good approximation to require that $a(\xi)$  continuously changes from $a_i$ to $a_{i+1}$ at $\xi_{i+1}$, which gives
\dis{\xi_0^i=\xi_{i+1}-\frac{1}{H_i}\sinh^{-1}\Big[\frac{H_i}{H_{i+1}}\sinh[H_{i+1}(\xi_{i+1}-\xi_0^{i+1})]\Big].}
In this way, every value of $\xi_0^i$ is determined by the value of $\xi_0^i$ in the region far outside the bubble.
Example of the profile of $a(\xi)$ is shown in Fig. \ref{Fig:AdSprof}.

     %%%%%%%%%%%%%%%%%%%%%%%%%%%%%%%%%%%%%%%%%%%%%%%%%%%%%%%%%%%%%%%%%%%%%%%%%%%%%%%%%%%%%%%%
%%%%%%%%%%%%%%%%%%%%%%%%%%%%%%%%%%%%%%%%%%%%%%%%%%%%%%%%%%%%%%%%%%%%%%

\begin{figure}[!t]
 \begin{center}
  \includegraphics[width=0.4\textwidth]{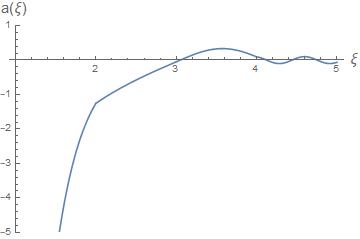}
 \end{center}
\caption{Profile of $a(\xi)$ describing the transition from dS to AdS space (at $\xi=3$), where  $\Lambda_4$ decreases discontinuously along the direction of $\xi \to 0$ : the values of $H$ are taken to be $3$ for $1<\xi<2$ (AdS), $1$ for $2<\xi<3$ (AdS), $3$ for $3<\xi<4$ (dS), and $10$ for $4<\xi<5$ (dS), respectively, and $\xi_0^4=0.2$.
  }
\label{Fig:dSprof}
\end{figure}

When the spacetime geometry before the bubble nucleation is given by   dS space,  $a_i(\xi)$ in the dS region (the region in which $\Lambda_4>0$) is  
\dis{a_i(\xi)=\frac{1}{H_i}\cos[H_i(\xi-\xi^{i}_0)].}
This shows that  in the dS region, $a(\xi)$ strongly oscillates for large value of $H$, but as we have seen in Sec. \ref{sec:6DEM},  $H$ is restricted to be smaller than $M_4 \lambda/(6\sqrt{\pi})$ at which $\frac{3f^2\lambda}{32\pi^2}=1$ is satisfied (see \eqref{eq:H2}) since otherwise the radion is not stabilized by the runaway behavior of the potential. 
This is remarkable because the obstruction of the radion stabilization is naturally connected to that of the semiclassical approximation : when $H>M_4$ the period of $a(\xi)$ is smaller than $M_4^{-1}$ but in this case $H$ is not fixed by the radion stabilization.
%We also note that as $\Lambda_4$ decreases toward the bubble of nothing, the value of $H$ decreases in the dS region, whereas it increases in the AdS region.
 For the transition between two dS spaces at $\xi_{i+1}$, the continuity between $a_i$ and $a_{i+1}$ leads to 
 \dis{\xi_0^i=\xi_{i+1}-\frac{1}{H_i}\cos^{-1}\Big[\frac{H_i}{H_{i+1}}\cos[H_{i+1}(\xi_{i+1}-\xi_0^{i+1})]\Big].}
 On the other hand, for the transition from dS   to AdS   space at $\xi =\xi_{i+1}$, the continuity condition becomes
 \dis{\xi_0^i=\xi_{i+1}-\frac{1}{H_i}\sinh^{-1}\Big[\frac{H_i}{H_{i+1}}\cos[H_{i+1}(\xi_{i+1}-\xi_0^{i+1})]\Big].}
 An example of the transition from dS to AdS space through the nucleation of the bubble is shown in Fig. \ref{Fig:dSprof}.
 
     %%%%%%%%%%%%%%%%%%%%%%%%%%%%%%%%%%%%%%%%%%%%%%%%%%%%%%%%%%%%%%%%%%%%%%%%%%%%%%%%%%%%%%%%
%%%%%%%%%%%%%%%%%%%%%%%%%%%%%%%%%%%%%%%%%%%%%%%%%%%%%%%%%%%%%%%%%%%%%%
  
As we have investigated, the transition between   vacua  through the bubble nucleation connects spacetime with any allowed value of $\Lambda_4$ to the nothing.
Indeed, since the nothing is the AdS vacuum with $\Lambda_4=-\infty$, the successful transition from the dS vacuum  to the nothing requires that the transition from the dS to the AdS or the Minkowski vacuum takes place without obstruction by the descent of the KK tower.
Regarding this, we recall that the descent of the KK tower becomes problematic when the combination $\frac{3 f^2\lambda}{32\pi^2}$ is kept fixed and  the limit  $\lambda \to 0$ is taken.
 In our case, in contrast, the transition  takes place by changing  $f$, while  $\lambda$ remains fixed so the transition from the dS vacuum to the nothing is not obstructed by the descent of the KK tower.
 If we can realize the UV completion which gives rise to the simultaneous changes in $f$ and $\lambda$ with $f^2 \lambda$ fixed, dS space may not be connected to the nothing through the bubble nucleation since $\Lambda_4$ does not change the sign under varying $\lambda$ and  the KK tower descends from UV in the limit $\lambda\to 0$.
 Then we  may say that  the realization of the metastable dS vacuum   is a counterexample of the cobordism conjecture or the metastable dS vacuum cannot exist according to the cobordism conjecture.

 \section{Conclusions}
\label{sec:conclusion}
 
The 6-dimensional Einstein-Maxwell theory compactified on $S^2$ contains essential ingredients of the moduli stabilization like the curvature of the extra dimensions, the flux through the extra dimensions, and the uplift, which also can be found in the string compactification.
 This motivates us to investigate the swampland conjectures concerning the moduli stabilization in the context of the 6-dimensional Einstein-Maxwell theory.
In this article, we first consider the scale separation and the scaling law, both of which are based on the distance conjecture.
We found that in two cases, the AdS vacuum with the vanishing uplift ($\lambda=0$) and any type of the vacuum with the fixed value of the combination $f^2\lambda$, the KK mass scale and the value of $\Lambda_4$ obey  the scaling law.
This indicates that, as claimed by the AdS/dS distance conjecture, (A)dS space with the tiny value of $\Lambda_4$ is not interpolated to the Minkowski space, but is   close to the swampland in which the KK tower descends from UV.
Moreover, the 6-dimensional Einstein-Maxwell theory allows the transition between  vacua through the nucleation of the bubble.
Of the particular interest is the case of $\Lambda_4=-\infty$, which was claimed to be identified with the `nothing'.
According to the cobordism conjecture, the background geometry consistent with  quantum gravity can smoothly evolve into the nothing at which the compact $S^2$ shrinks to zero size.
 Since the combination $f^2\lambda$ is not fixed in this case, that is, the flux is changed by the bubble wall but the uplift remains fixed, the transition from the metastable dS vacuum to the nothing is not obstructed by the descent of the KK tower.
 This implies that the metastable dS vacuum in the 6-dimensional Einstein-Maxwell is not excluded by the cobordism conjecture.
  
  It is quite remarkable that whether the given vacuum belongs to the landscape or the swampland crucially depends on the relation between parameters representing different ingredients of the moduli (radion in this case) potential.
  That is, if the flux $f$ and the uplift $\lambda$ are independent, the metastable dS vacuum is not obstructed by the descent of the KK tower and is cobordant to the nothing.
  Then the distance and the cobordism conjecture cannot be used to claim that the metastable dS vacuum belongs to the swampland.
  In contrast, the opposite conclusion can be drawn if $f$ and $\lambda$ are no longer independent, but constrained by $f^2 \lambda$=(constant). 
  Since such a relation can be explained in the UV completion,  this may suggest that the EFT alone is not sufficient to distinguish the landscape from the swampland, but the appropriate UV completion needs to be taken into account.

\subsection*{Acknowledgements} 
%

%

%\newpage

\appendix

\renewcommand{\theequation}{\Alph{section}.\arabic{equation}}

\section{Details of the bubble solution }
\label{app:wholesol}
\setcounter{equation}{0}
 
 In the thin wall approximation, the value of $\Lambda_4$  inside (or outside) the bubble  can be approximated by a constant unless the narrow region near the bubble wall is considered.
 Thus, the geometry of the region inside (or outside) the bubble   is well described by   Euclidean (A)dS or Minkowski space, depending on the sign of $\Lambda_4$.
 Meanwhile, at the  bubble wall, the black 2-brane which changes the flux, hence   $\Lambda_4$, is located.
Then even though $a(\xi)$ and $\phi(\xi)$ are continuous at the bubble wall, their derivatives are not, as studied in \cite{Brown:2010mf}.
To see this in detail, we consider the action describing the whole region, i.e., the bubble wall as well as the regions   inside  and outside the bubble   :
\dis{S_E=&S_E^{\rm in}+S_E^{\rm brane}+S_E^{\rm out}
\\
=&2\pi^2\int_0^{\overline \xi} d\xi\Big[a^3\Big(\frac12{\phi'}^2+V\Big)+3M_4^2\Big(\frac{d}{d\xi}(a^2a')-a {a'}^2-a\Big)\Big]
\\
&+2\pi^2 a^3 Te^{-\frac32\frac{\phi}{M_4}}\Big|_{\overline \xi}
\\
&+2\pi^2\int_{\overline \xi}^{\infty} d\xi\Big[a^3\Big(\frac12{\phi'}^2+V\Big)+3M_4^2\Big(\frac{d}{d\xi}(a^2a')-a {a'}^2-a\Big)\Big],
}
where $\overline{\xi}$ is the location of the bubble wall, and  we included the DBI action for the black 2-brane.
The brane tension is given by $T=\frac{2}{\sqrt3} (ng_6) M_6^2$ ($n$: the number of branes), which is identified with the ADM mass of the brane solution \eqref{eq:brane} (see Eq. (2.6) of \cite{Lu:1993vt}).
Under the variations $\delta a$ and $\delta \phi$, $\delta S_E^{\rm in}$ and $\delta S_E^{\rm out}$ provide  the surface term containing $a'$ and $\phi'$.
That is, under  $\delta \phi$ the kinetic term gives $2\pi^2  a^3\phi'\delta \phi $.
Meanwhile, under $\delta a$ the Einstein-Hilbert term gives $12\pi^2 M_4^2 a a'\delta a+12\pi^2M_4^2 a^2\delta a'= \frac{d}{d\xi}(6\pi^2 M_4^2 a^2\delta a)$ from the first surface term  and $-12\pi^2M_4^2 a a'\delta a$ from the second term, respectively.
The former, the surface term of the surface term is defined on the boundary of the boundary which is empty, so we may ignore it.
Identifying the sum of  these variations  with the variation of the brane term, $(6\pi^2a^2\delta a -3\pi^2 a^3\frac{\delta \phi}{M_4})Te^{-\frac32 \frac{\phi}{M_4}}$, we obtain
\dis{&\Delta \phi'\equiv \phi'_{\rm out}-\phi'_{\rm in}=-\frac32 \frac{T}{M_4}e^{-\frac32 \frac{\phi}{M_4}},
\\
&\Delta a'\equiv a'_{\rm out}-a'_{\rm in}=-\frac{T}{2M^2}ae^{-\frac32 \frac{\phi}{M_4}}. }  
In particular, the second relation gives  
\dis{\Delta \Big(\frac12 {\phi'}^2-V\Big)&=M_4^2\Delta \frac{3{a'}^2-1}{a^2}=M_4^2\frac{6a'}{a}\Delta a
\\
&=-3T\frac{a'}{a}e^{-\frac32 \frac{\phi}{M_4}},}
where $a'$ can be taken to be the mean value between $a'({\overline \xi}+\epsilon)$ and $a'({\overline \xi}-\epsilon)$ ($\epsilon \ll 1$).

On the other hand, the transition between vacua through the nucleation of the bubble includes the `up tunnelling', in which the value of $\Lambda_4$ inside the bubble  is larger than that outside the bubble.
The qualitative feature of  the  up tunnelling rate crucially depends on the sign of $\Lambda_4$ outside the bubble.
 In particular, if the geometry outside the bubble   is AdS or Minkowski space, the Euclidean version of which is noncompact, the up tunnelling rate vanishes \cite{Brown:2011gt}.
 To see this, consider two vacua, say, A vacuum and B vacuum where the value of $\Lambda_4$ in A vacuum is larger than that in  B vacuum.
Then the tunnelling rate from  A vacuum to  B vacuum satisfies \cite{Coleman:1980aw}
\dis{\Gamma_{A\to B} \sim e^{-(S_E({\rm instanton})-S_E(A))}.}
At the same time, there can be the `up tunnelling', the transition from  B vacuum to  A vacuum, the rate of which is given by
\dis{\Gamma_{B\to A}\sim e^{-(S_E({\rm instanton})-S_E(B))}.}
Here the instanton action $S_E$(instanton) in $\Gamma_{A\to B}$ is the same as that in $\Gamma_{B\to A}$ and finite.
Now, $S_E(A)$ and $S_E(B)$ can be obtained by putting the solutions to the equations of motion \eqref{eq:EoM} into \eqref{eq:action1}.
Replacing ${\phi'}^2/2$ by $V+3M_4^2[({a'}^2-1)/a^2]$, we obtain \cite{Eckerle:2020opg}
\dis{S_E^{\rm in}=4\pi^2 \int d\xi (a^3V-3 M_4^2 a)+6\pi^2M_4^2 \int d\xi\frac{d}{d\xi}(a^2a'),\label{eq:actionEoM}}
where the last surface term will not be considered since we are interested in the behaviors of $a(\xi)$ and $\phi(\xi)$ in the bulk.
The first equation of motion in \eqref{eq:EoM} gives the relation
\dis{da =d\xi \sqrt{1-\frac{V}{3 M_4^2}a^2},}
from which the first integral in \eqref{eq:actionEoM} becomes
\dis{S_E^{\rm in}=-12\pi^2 M_4^2\int da a\sqrt{1-\frac{V}{3M_4^2}a^2}=12\pi^2\frac{M_4^2}{V}\Big(1-\frac{V}{3M_4^2}a^2\Big)^{3/2}\Big|_{a_i}^{a_f},}
where the surface term is omitted.
Moreover, $S_E^{\rm out}$, the action for the region outside the bubble  is also given by the same expression.
Therefore, if the geometry of the region outside the bubble   is AdS space, $V<0$ and $a_f$ extends to infinity, thus $S_E^{\rm out}({\rm AdS})=-\infty$.
Then the ratio of the transition rates $\Gamma_{B\to A}/\Gamma_{A\to B}=e^{S_E(B)-S_E(A)}$ becomes zero if the geometry of B vacuum (outside the bubble wall) is AdS space.
The same conclusion can be drawn when  the geometry of B vacuum is Minkowski space, in which $V=0$ and $a=\xi$, thus $S_E^{\rm out}=-6\pi^2 M_4^2 (a_f^2-a_i^2)$ with $a_f=\infty$.
This shows that the nucleation of the bubble of the lager value of $\Lambda_4$ in AdS or Minkowski space is forbidden.

\end{document}